\documentclass{elsart}

\usepackage{natbib}

\usepackage{graphicx}
\usepackage{epsfig}

\usepackage{amssymb}

\begin{document}

\begin{frontmatter}



\title{\bf Exploring the Structure of Galaxy Clusters: 
XMM-Newton observations of the REFLEX-DXL clusters at $z\sim 0.3$
\thanksref{titlefn}}
\thanks[titlefn]{This work is based on observations
made with the XMM-Newton, an ESA science mission with 
instruments and contributions directly funded by
ESA member states and the USA (NASA).}



\author[MPE]{Y.-Y. Zhang},
\author[MPE]{H. B\"ohringer}, 
\author[MPE]{A. Finoguenov}, 
\author[MPE,BAT]{Y. Ikebe}, 
\author[MPE,TOK]{K. Matsushita}, 
\author[MPE]{P. Schuecker},
\author[INA]{L. Guzzo},
\author[LIV]{C. A. Collins}
\address[MPE]{Max-Planck-Institut f\"ur extraterrestrische Physik, Garching, Germany}
\address[BAT]{Joint Center for Astrophysics, University of Maryland, Baltimore, USA}
\address[TOK]{Tokyo University of Science, Tokyo, Japan}
\address[INA]{INAF-Osservatorio Astronomico di Brera, Merate/Milano, Italy}
\address[LIV]{Liverpool John Moores University, Liverpool, U.K.}

\begin{abstract}

The precise determination of global properties of galaxy clusters, and
their scaling relations, is a task of prime importance for the use of
clusters as cosmological probes. We performed a detailed XMM-Newton
study of 14 X-ray luminous REFLEX Survey clusters at $z \sim 0.3$. We
found that the properties of the galaxy clusters show a self-similar
behavior at $r>0.1 r_{\rm vir}$. This helps to establish tighter
scaling relations. Peculiarities in the individual clusters are
important to understand the scatter from the self-similar frame in the
cluster central parts.

\end{abstract}


\end{frontmatter}


\section{Introduction}
\label{intro}

ROSAT and ASCA observations (e.g. Markevitch et al. 1998; Vikhlinin et
al. 1999; Arnaud et al. 2002; Reiprich and B\"ohringer 2002) and
simulations (e.g. Kay 2004; Borgani 2004) indicate a self-similar form
of the intracluster medium (ICM) properties such as temperature,
density, and entropy for massive clusters ($k_{\rm B}T > 4$~keV)
excluding cooling cores (Fabian and Nulsen 1977). XMM-Newton has the
advantage of high spectral resolution and large field of view (FOV)
for detailed studies. This helps us to compose an almost volume
complete sample of 13 distant, X-ray luminous (DXL; $z=0.27$ to
$0.31$; $L_{X}~\geq~10^{45}~{\rm erg~s^{-1}}$ for $0.1-2.4~{\rm keV}$)
galaxy clusters and one supplementary cluster at $z=0.2578$ from the
ROSAT-ESO Flux-Limited X-ray (REFLEX; B\"ohringer et al. 2004) galaxy
cluster survey. We correct the volume completeness with a known
selection function for distant, X-ray luminous clusters
($L_{X}~\geq~10^{45}~{\rm erg~s^{-1}}$) as described in B\"ohringer et
al.  (2005; Paper~I).  A prime goal for the study of the REFLEX-DXL
sample is to obtain spatially resolved ICM properties such as the
temperature (Zhang et al. 2004a; Paper\,II), to derive accurate
measurements of the cluster mass and gas mass fraction, and to study
the peculiarities in the cluster structure which introduce a scatter
in the scaling relations.  We adopt a flat $\Lambda$CDM cosmology with
$\Omega_{\rm m}=0.3$ and $H_{\rm 0}=70$~km~s$^{-1}$~Mpc$^{-1}$.
Confidence intervals correspond to the 68\% confidence level.

\begin{figure}
\includegraphics[width=10cm,angle=270,clip=]{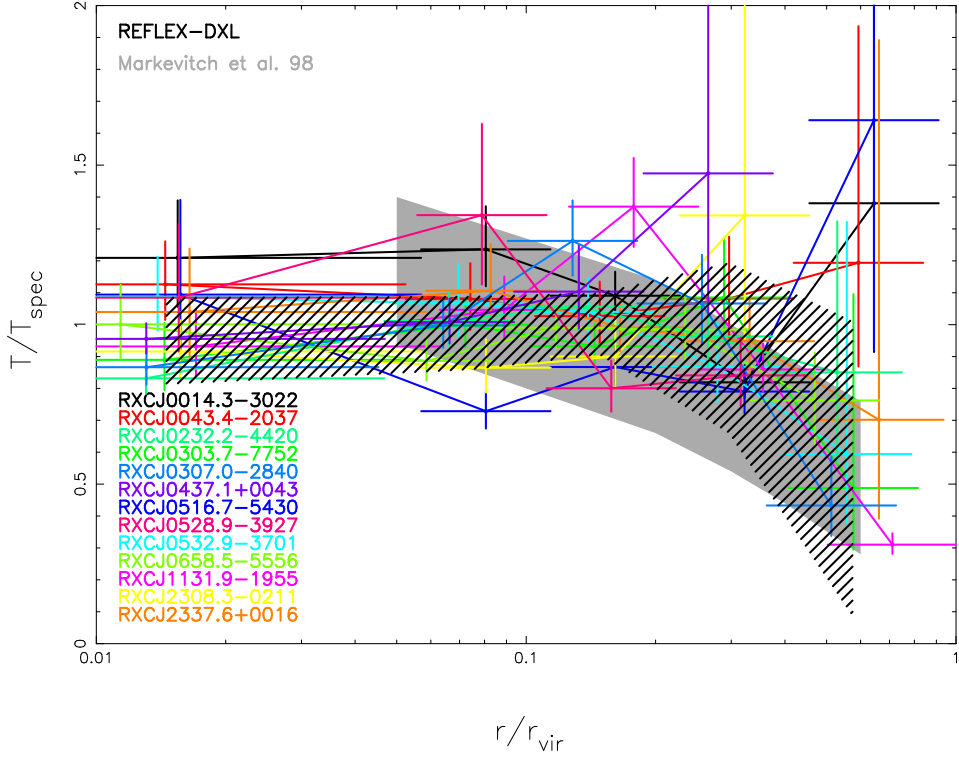}
\label{figure1}
\caption{Scaled temperature profiles.
The shadow shows a weighted temperature profile of the
REFLEX-DXL clusters (hatched) and the temperature profile range
in Markevitch et al. (1998; filled). }
\end{figure}

\section{Data reduction}
\label{s:method}

\begin{figure}
\includegraphics[width=10cm,angle=270,clip=]{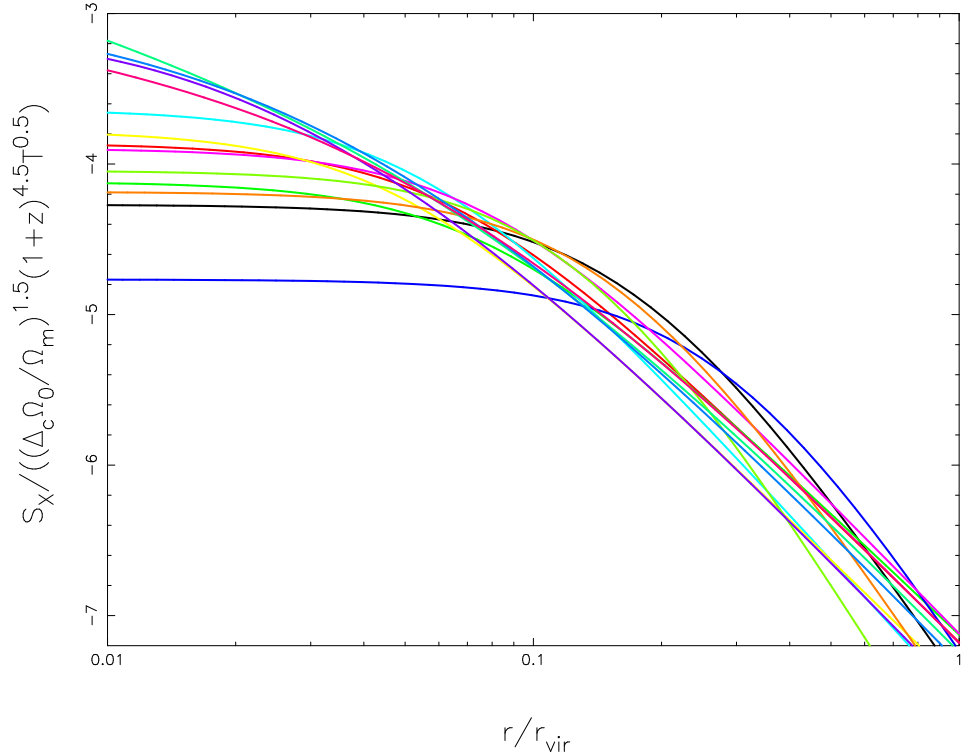}
\label{figure2}
\caption{
The best surface brightness model fits (PSF deconvolved) scaled to the
self-similar frame for pn. The color coding is the same as used for
Fig.~\ref{figure1}. The typical uncertainty of the surface brightness
increases from 10\% to 20\% from the inner parts to the outer parts.}
\end{figure}

We use the XMMSAS v6.0 software for data reduction. For pn data, the
fractions of the out-of-time (OOT) effect are 2.32\% and 6.30\% for
Extend Full Frame (EFF) and Full Frame (FF) mode, respectively. We
create an OOT event list file and statistically remove the OOT effect.
To avoid the episodes of ``soft proton flares'' (De
Luca \& Molendi 2004), we use a threshold of $3 \sigma$ clipping to
clean the data in both the hard band (10--12~keV band and 12--14~keV
band for MOS and pn, respectively) and the soft band (0.3--10~keV
band).

Over half of the clusters clearly show substructures or/and
elongation. We use ``edetect\_chain'' to detect point-like sources.
We subtract the substructures and point-like sources leaving only the
main component.

The observations of RXCJ2011.3$-$5725 are contaminated by flares. Thus
we only obtain a global temperature ($\sim 3.77$~keV). Some properties
of these observations and an overview of the sample are described in
Paper\,I.  We obtained the temperature profiles of 9 REFLEX-DXL
clusters in Paper\,II. Also included in this paper are the
observational parameters, alternative names, data preparation and
double background subtraction method which is developed to provide a
precise spectral background removal. We apply the XMM-Newton blank sky
pointings in the Chandra Deep Field South (CDFS) as background.

\section{ICM properties}
\label{s:result}

We have considered the projection effects over the line of sight in
the temperature and surface brightness measurements. We have accounted
for the PSF effect for surface brightness. In the spectral analysis,
we have applied a large radial binning, greater than $0.5^{\prime}$,
to reduce the PSF effect instead of accounted for the PSF blurring
completely because of limited photon statistics. This should be
considered for the reliability of the temperature profiles in the
central part. For such distant clusters, the PSF effect is important
within $0.3r_{\rm vir}$ which introduces an added uncertainty to the
final results. This can only be investigated using deep exposures with
better photon statistics.

\begin{figure}
\includegraphics[width=10cm,angle=270,clip=]{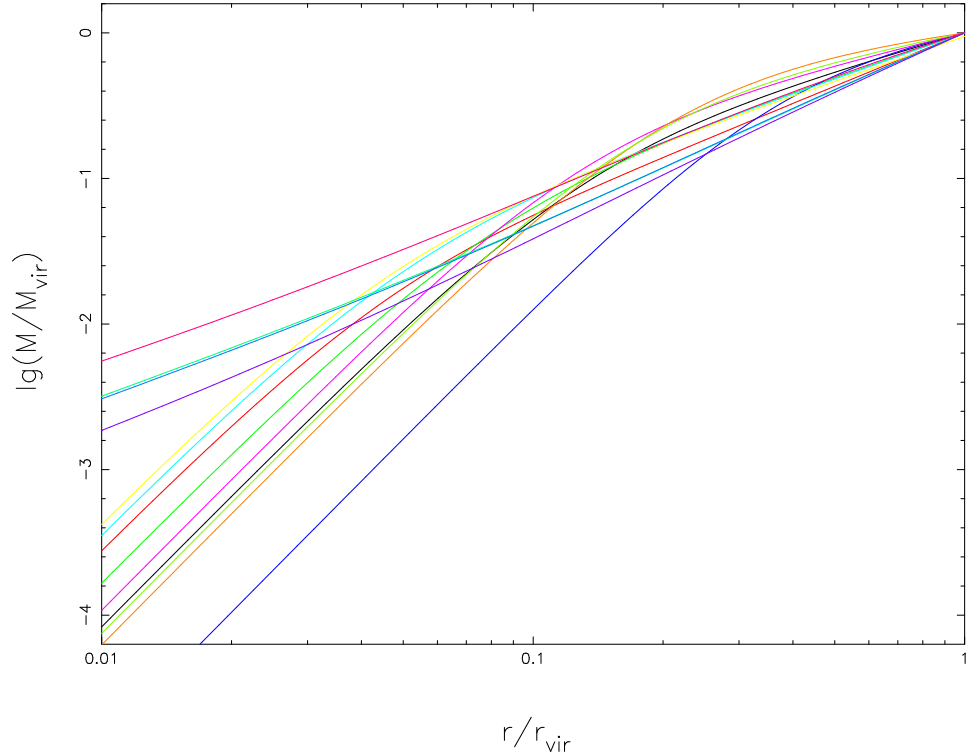}
\label{figure3}
\caption{Scaled X-ray mass profiles of the REFLEX-DXL clusters. 
The color coding is the same as used for Fig.~\ref{figure1}. The
typical uncertainty of the mass increases from 10--40\% to 25--80\%
from the inner parts to the outer parts.}
\end{figure}

\subsection{Temperature distribution} 

We perform the spectral analysis in five annuli: 0--0.5$^\prime$,
0.5--1$^\prime$, 1--2$^\prime$, 2--4$^\prime$, and 4--8$^\prime$, to
measure the temperature profiles. We applied both the residual
spectrum (e.g. Arnaud et al. 2002; Zhang et al. 2005) and a
``powerlaw'' model (Zhang et al. 2004a) to account for the residual
background in the double background subtraction method (Zhang et
al. 2004a) for spectral background removal. We found that the two methods
provide consistent results within 1~$\sigma$ error except for the last
annulus for RXCJ0232.2$-$4420 and RXCJ1131.9$-$1955. The disagreement
is caused by the underestimate/overestimate of the formal errors in
the measurements using a residual model/spectrum. We deprojected the
temperature profiles and found consistent results within 1~$\sigma$
error bars.

We apply the global spectral temperatures and $r_{\rm vir}$ to scale
the temperature profiles (see Fig.~\ref{figure1}). Excluding the
individual data points with error bars over 150\% of the mean value, we
derived a weighted temperature profile of the REFLEX-DXL clusters. We
found a closely self-similar behavior with a constant distribution at
$r<0.3 r_{\rm vir}$ and a decrease at $r>0.3 r_{\rm vir}$. This
averaged temperature profile keeps an overall agreement with the one
in Markevitch et al. (1998).  A similar universal temperature profile
is indicated in simulations (Borgani et al. 2004; Borgani 2004).

\subsection{Surface brightness}
\label{s:sx}

We choose the 0.5--2~keV band to derive surface brightness profiles
(also see Zhang et al. 2005). This provides an almost
temperature-independent X-ray emission coefficient over the expected
temperature range. We derive the azimuthally averaged surface
brightness of the XMM-Newton blank sky pointings in the same detector
coordinates as for the targets. The count rate ratio of the targets
and CDFS in the 10--12~keV band and 12--14~keV band for MOS and pn,
respectively is used to scale the CDFS surface brightness. The data
are rebinned to ensure (1) at leat 100 counts (30 counts for a limited
photon statistics case) per bin, and (2) 3-$\sigma$ source count
rate. We subtract the scaled CDFS surface brightness and obtain the
surface brightness including the cluster surface brightness and a
residual soft X-ray background. This residual background is flat over
the FOV. It is estimated in the outer region
($11^{\prime}<r<15^{\prime}$). We fit the cluster surface brightness
profile by a surface brightness prediction convolved with the
XMM-Newton Point Spread Function (PSF; Ghizzardi 2001).

Four (RXCJ0232.2$-$4420, RXCJ0307.0$-$2840, RXCJ0437.1$+$0043 and
RXCJ0528.9$-$3927) of 13 clusters show pronounced or moderate cooling
flows.  For the remaining 9 clusters, a $\beta$-model provides a
satisfying $\chi^2$ fit of the surface brightness profiles in the
observed radial range. The surface brightness profiles (see
Fig.~\ref{figure2}) are scaled according to the standard
self-similar model (e.g. Arnaud et al. 2002) and show a good
self-similarity in the $r>0.1 r_{\rm vir}$ region.

\subsection{Mass distribution}

Surface brightness profiles can be deprojected to yield the emission
per volume element, $\xi (r)=\widetilde{\Lambda}(r) n^2_{\rm
e}(r)$. This is used to derive ICM density distributions.

For Non-Cooling flow Clusters (NCCs), we make use of the temperature
profile and density profile to derive the mass, 
assuming spherical symmetry and hydrostatic equilibrium. The
XMM-Newton mass measurements of the NCCs are consistent with the
masses derived from the global temperatures using the observational
$M_{200}$--$T$ relation based on the conventional $\beta$ model for
the X-ray surface brightness profile and hydrostatic equilibrium for
22 nearby clusters from Xu et al. (2001).

Cooling flow Clusters (CCs) show a surface brightness excess in the
center. Navarro et al. (1997; $\alpha=1$; NFW) describe a universal
density profile for dark halos from numerical simulations in
hierarchical clustering scenarios. This fits the steep X-ray mass
profile of CCs in the central region to some degree.  An extended-NFW
model from recent simulations (e.g. Diemand et al. 2004; Navarro et
al. 2004; $\alpha>1$; ext-NFW) provides an improved fit for such a
cuspy profile in the cluster center.  For CCs, we make use of the
deprojected temperature profile and density profile to parameterize
the ext-NFW model under the assumptions of hydrostatic equilibrium,
polytropic gas and spherical symmetry. Then the mass is derived from
the ext-NFW model.

The typical uncertainty of the mass increases from 10--40\% to
25--80\% from the inner parts to the outer parts. The
NCCs/CCs show shallow/steep mass profiles at the inner parts
(Fig.~\ref{figure3}). We extrapolate the mass distributions beyond
the observed radial range up to the virial radii and scale the mass
profiles using the virial mass. The scaled mass profiles
(Fig.~\ref{figure3}) show a self-similar behavior in the $r>0.1
r_{\rm vir}$ region. The mass concentrations are different for NCCs
and CCs in the center.

\subsection{Gas mass fraction distribution}

\begin{figure}
\includegraphics[width=10cm,angle=270,clip=]{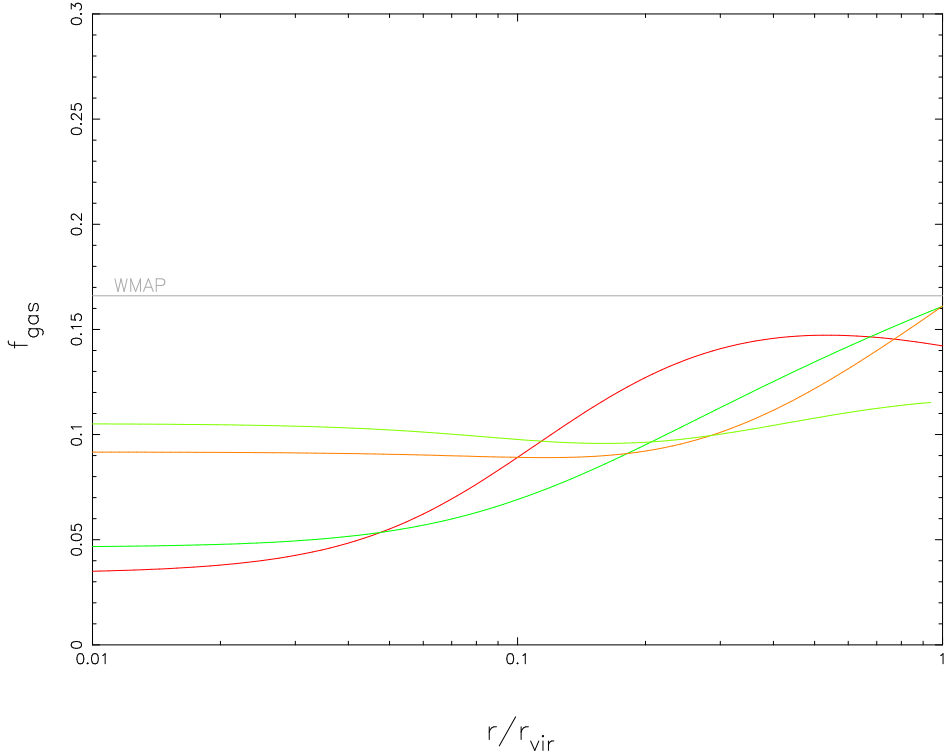}
\label{figure4}
\caption{Gas mass fraction profiles for four typical examples.
The WMAP measured baryon fraction of the Universe is $f_{\rm
b}=0.166$, where $\Omega_{\rm b}~ h^2=0.0224$ and $\Omega_{\rm m}~
h^2=0.135$ (Spergel et al. 2003, horizontal line). The color coding is
the same as used for Fig.~\ref{figure1}. The typical uncertainty is in
the range of 20--70\% at $r_{\rm vir}$.}
\end{figure}

The gas mass fraction (see Fig.~\ref{figure4}) as a function of
radius is derived by $f_{\rm gas}(r)=M_{\rm gas}(r)/M(r)$. At
$r_{2500}$, the gas mass fractions derived from the XMM-Newton
exposure of around 10~ks agree with the measurements of Allen et
al. (2004) based on the Chandra observations of 7 clusters yielding
$f_{\rm gas}\sim 0.105$--$0.138h^{-3/2}_{70}$ with similar confidence
intervals ($\sim 0.02h^{-3/2}_{70}$). The gas mass fraction
distributions between $r_{500}$ and $r_{200}$ are in the range of
$0.12 \sim 0.24$. This is in good agreement with the WMAP measurement
of 0.166 (Spergel et al. 2003), the measurements of Ettori et
al. (2002) which are based on BeppoSAX observations of 22 nearby
clusters, and the measurements of Sanderson et al. (2003) which are
based on ASCA/GIS, ASCA/SIS and ROSAT/PSPC observations of 66 clusters
yielding $f_{\rm gas}=0.13\pm 0.01~h_{70}^{-3/2}$ at
$r_{200}$. RXCJ0014.3$-$3022, RXCJ0516.7$-$5430 and RXCJ1131.9$-$1955
are in the stage of merger. We obtain slightly higher gas mass
fractions in the outskirts artificially caused by the assumption of
spherical symmetry.

\section{Summary and Discussion}
\label{s:conclusion}

We derive the spatially resolved temperature profiles.  We combine the
temperature profile with the gas density profile observed up to almost
$r_{500}$ to compute accurately the mass and gas mass fraction.  An
ext-NFW model ($\beta$-model) provides a good fit for the CCs
(NCCs). X-ray luminous (massive) galaxy clusters such as the clusters
in this paper show a self-similar behavior of their properties,
e.g. temperature, density and mass in the $r>0.1 r_{\rm vir}$ region
in which we have a good understanding of dark matter.

Peculiarities in the cluster structure introduce a scatter from the
self-similar frame in the cluster cores which is due to physical
processes rather than simply being statistical fluctuations in the
measurement (Zhang et al. 2004b). For example, mergers can not only
lead to a temporary increase in the cluster temperature and X-ray
luminosity (Randall et al. 2002), but also reduce the surface
brightness slope and increase the core radius particular in on-going
cluster mergers (Markevitch et al. 2002), e.g. RXCJ0658.5$-$5556 and
RXCJ0516.7$-$5430 in our sample.  Many studies (e.g. Markevitch et
al. 2002; Randall et al. 2002) indicate that the X-ray mass estimate
in the center could be biased by the phenomena such as ghost cavities
and bubbles that may somehow invalidate the hydrostatic equilibrium
hypothesis.  More details of the REFLEX-DXL cluster properties,
scaling relations, correlations and their scatters will be described
in forthcoming papers.

\section*{Acknowledgments} 

YYZ acknowledges receiving the International Max-Planck Research
School Fellowship. YYZ would like to thank Hermann Brunner, Michael
Freyberg, and Rasmus Voss for providing useful suggestions.

%


\begin{thebibliography}{}

\bibitem[N(1999)]{a} Allen, S. W., Schmidt, R. W., Ebeling, H. et al. Constraints on dark energy from Chandra observations of the largest relaxed galaxy clusters. MNRAS 353, 457-467, 2004. 
\bibitem[N(1999)]{a} Arnaud, M., Aghanim, N., Neumann, M. The X-ray surface brightness profiles of hot galaxy clusters up to vec $z \sim 0.8$: Evidence for self-similarity and constraints on $\Omega_{0}$. A\&A 389, 1-18, 2002.
\bibitem[N(1999)]{a} Borgani, S. Galaxy clusters as cosmological tools and astrophysical laboratories. Proc. The Riddle of Cooling Flows in Galaxies and Clusters of Galaxies: E4., ed. T. H. Reiprich, J. C. Kempner, \& N. Soker, http://www.astro.virginia.edu/coolflow/proc.php?regID=115, 2004.
\bibitem[N(1999)]{a} Borgani, S., Murante, G., Springel, V. et al. X-ray properties of galaxy clusters and groups from a cosmological hydrodynamical simulation. MNRAS 348, 1078-1096, 2004.
\bibitem[N(1999)]{a} B\"ohringer, H., Schuecker, P., Guzzo, L. et al. The ROSAT-ESO flux limited X-ray (REFLEX) galaxy cluster survey. V. The cluster catalogue. A\&A 425, 367-383, 2004.
\bibitem[N(1999)]{a} B\"ohringer, H., Schuecker, P., Zhang, Y.-Y. et al. The distant, X-ray luminous galaxy cluster sample from the REFLEX survey (REFLEX-DXL) and its temperature function. A\&A submitted, 2005.
\bibitem[N(1999)]{a} De Luca, A., \& Molendi, S. The 2-8 keV cosmic X-ray background spectrum as observed with XMM-Newton. A\&A 419, 837-848, 2004.
\bibitem[N(1999)]{a} Diemand, J., Moore, B., \& Stadel, J. Convergence and scatter of cluster density profiles. MNRAS 353, 624-632, 2004.
\bibitem[N(1999)]{a} Ettori, S., De Grandi, S., Molendi, S. Gravitating mass profiles of nearby galaxy clusters and relations with X-ray gas temperature, luminosity and mass. A\&A 391, 841-855, 2002.
\bibitem[N(1999)]{a} Fabian, A. C., Nulsen, P. E. J. Subsonic accretion of cooling gas in clusters of galaxies. MNRAS 180, 479-484, 1977.
\bibitem[N(1999)]{a} Ghizzardi, S. In-flight calibration of the PSF for the MOS1 and MOS2 cameras. EPIC-MCT-TN-011 (Internal report), 2001.  
\bibitem[N(1999)]{a} Kay, S. T. The entropy distribution in clusters: Evidence of feedback? MNRAS 347, L13-L17, 2004.
\bibitem[N(1999)]{a} Markevitch, M., Forman, W., Sarazin, C. L. et al. The temperature structure of 30 nearby clusters observed with ASCA: Similarity of temperature profiles. ApJ 503, 77-96, 1998.
\bibitem[N(1999)]{a} Markevitch, M., Gonzalez, A. H., David, L. et al. A Textbook Example of a Bow Shock in the Merging Galaxy Cluster 1E 0657$-$56. ApJ 567, L27-L31, 2002.
\bibitem[N(1999)]{a} Moore, B., Quinn, T., Governato, F. et al. Cold collapse and the core catastrophe. MNRAS 310, 1147-1152, 1999.
\bibitem[N(1999)]{a} Navarro, J. F., Frenk, C. S., White, S. D. M. A universal density profile from hierarchical clusterin. ApJ 490, 493-508 (NFW), 1997. 
\bibitem[N(1999)]{a} Navarro, J. F., Hayashi, E., Power, C. et al. The inner structure of LambdaCDM haloes - III. Universality and asymptotic slopes. MNRAS 349, 1039-1051 (ext-NFW), 2004. 
\bibitem[N(1999)]{a} Randall, S. W., Sarazin, C. L., Ricker, P. M. The Effects of Mergers Boosts on the luminosity, temperature and inferred mass functions of clusters of galaxies. ApJ 577, 579-594, 2002.
\bibitem[N(1999)]{a} Reiprich, T. H. \& B\"ohringer, H. The mass function of an X-ray flux-limited sample of galaxy clusters. ApJ 567, 716-740, 2002.
\bibitem[N(1999)]{a} Sanderson, A. J. R., Ponman, T. J., Finoguenov, A. et al. The Birmingham-CfA cluster scaling project - I. Gas fraction and the $M$--$T_{X}$ relation. MNRAS 340, 989-1010, 2003.
\bibitem[N(1999)]{a} Spergel, D. N., Verde, L., Peiris, H. V. et al. First-year Wilkinson Microwave Anisotropy Probe (WMAP) observations: Determination of cosmological parameter. ApJS 148, 175-194, 2003.
\bibitem[N(1999)]{a} Vikhlinin, A., Forman, W., Jones, C. Outer regions of the cluster gaseous atmospheres. ApJ 525, 47-57, 1999.
\bibitem[N(1999)]{a} Xu, H.-G., Jin, G.-X., Wu, X.-P. The mass-temperature relation of 22 nearby clusters. ApJ 553, 78-83, 2001.
\bibitem[N(1999)]{a} Zhang, Y.-Y., Finoguenov, A., B\"ohringer, H. et al. Temperature gradients in XMM-Newton observed REFLEX-DXL galaxy clusters at $z \sim 0.3$. A\&A 413, 49-63, 2004a.
\bibitem[N(1999)]{a} Zhang, Y.-Y., Finoguenov, A., B\"ohringer, H. et al. Spatial distributions of the REFLEX-DXL galaxy clusters at $z \sim 0.3$ observed by XMM-Newton. Proc. Memorie della Societb Astronomica Italiana - Supplementi in press, astro-ph/0402533, 2004b.
\bibitem[N(1999)]{a} Zhang, Y.-Y., B\"ohringer, H., Mellier, Y. et al. XMM-Newton study of the lensing cluster of galaxies CL0024$+$17. A\&A 429, 85-99, 2005.
\end{thebibliography}
\end{document}